\documentstyle[12pt,twoside,fleqn,espcrc1,epsfig]{article}

\newcommand{\AmS}{{\protect\the\textfont2
  A\kern-.1667em\lower.5ex\hbox{M}\kern-.125emS}}

\title{$\pi^0\pi^0$ Scattering Amplitudes and Phase Shifts 
Obtained
by the $\pi^-p$ Charge Exchange Process}

\author{Kunio Takamatsu\address{Miyazaki U., Gakuen-Kibanadai, 
Miyazaki 889-2155, Japan\\ 
Presented for the collaboration(A.M.Ma, K. Takamatsu, 
M.Y.Ishida, S.Ishida, T.Ishida, 
T. Tsuru and H. Shimizu) and the E135 collaboration.
}}
       
\begin{document}

\maketitle

\begin{abstract}
The results of the analysis of the $\pi^0\pi^0$ scattering amplitudes 
obtained with
$\pi^-p$ charge exchange reaction, 
$\pi^-p\rightarrow \pi^0\pi^0n$, data at 9 GeV/c
are presented. The $\pi^0\pi^0$ 
scattering amplitudes show clear $f_0(1370)$ and $f_2(1270)$ 
signals in the S and D waves, 
respectively.
The  $\pi^0\pi^0$  scattering 
phase shifts have been 
obtained below $K\bar K$ threshold and been analyzed 
by the Interfering Amplitude 
method with introduction of negative background phases. 
The results show a S wave 
resonance, $\sigma$. Its 
Breit-Wigner parameters are in good agreement with those of our 
previous analysis on the $\pi^+\pi^-$ phase shift data.
\end{abstract}

~

The fact that no odd wave is present in $\pi^0\pi^0$ scattering amplitude
in comparison with $\pi^+\pi^-$ is an advantage. 
However experimental difficulties impaired up to now the 
quality of $\pi^0\pi^0$ data \cite{rf1} for the analysis of
$\pi\pi$ phase shifts.
Only the high statistics $\pi^0\pi^0$ data of Cason et al.\cite{rf2} 
have been used for analysis 
so far, but they behave differently from 
the $\pi^+\pi^-$\cite{rfCM} data 
below $K\bar K$ threshold 
and cannot help to solve the ambiguity of the solutions in the 
analysis of $\pi^+\pi^-$ data.

	We have analyzed the  $\pi^0\pi^0$  system produced in 
the $\pi^-p$ charge exchange 
process $\pi^-p\rightarrow \pi^0\pi^0 n$ at 9 GeV/c studied by the 
E135 experiment with the Benkei 
spectrometer\cite{rf3} at the KEK12 GeV PS. 
We have obtained the  $\pi^0\pi^0$  scattering amplitudes 
and also the $\pi^0\pi^0$  scattering phase shifts below $K\bar K$ 
threshold. 

	Fig 1 shows the acceptance corrected  $\pi^0\pi^0$  mass 
distribution reconstructed from 
$4\gamma$'s in the final state. 
The off-mass shell scattering amplitudes 
$T_{\pi\pi}(m_{\pi\pi}^2, {\rm cos}\theta , t)$ are 
extrapolated to the on-mass shell scattering amplitudes 
at the pion pole, 
$T_{\pi\pi} (m_{\pi\pi}^2, {\rm cos}\theta , m_\pi^2)$, 
as the process proceeds through one pion exchange. 
A linear extrapolation is adopted. 
The on-mass shell scattering amplitudes can be described, 
considering the S and D waves, as follows; 
$T_{\pi\pi} (m_{\pi\pi}^2, {\rm cos}\theta , m_\pi^2) 
= A_S + A_D 5(3{\rm cos}^2\theta -1)/2$, where $A_S$ and 
$A_D$ are S and D wave scattering amplitudes, respectively.

\begin{figure}
  \epsfysize=6.5 cm
 \centerline{\epsffile{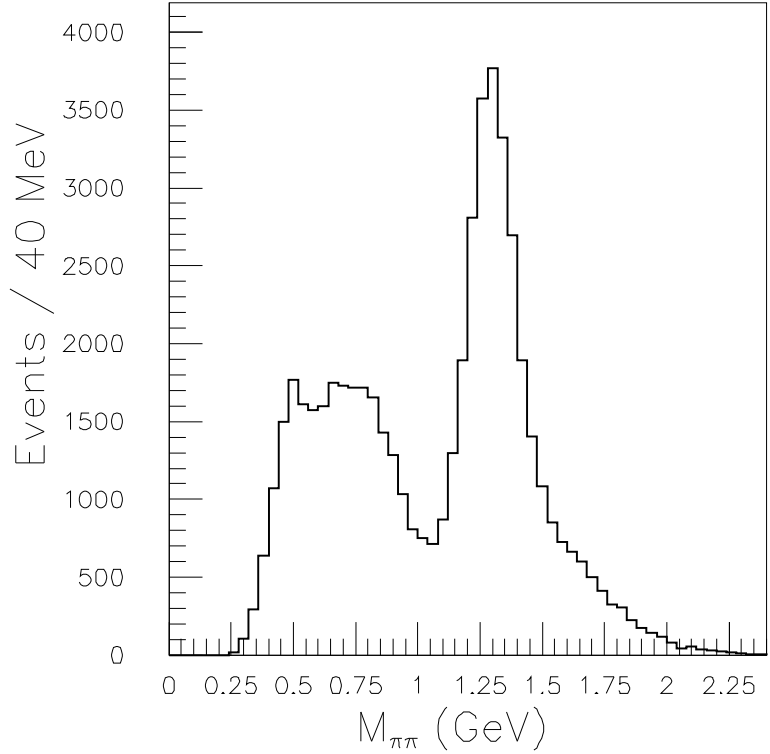}}
 \caption{ Acceptance corrected  $\pi^0\pi^0$ mass distribution}
  \label{fig1}
\end{figure}

\begin{figure}
  \epsfysize=6.5 cm
 \centerline{\epsffile{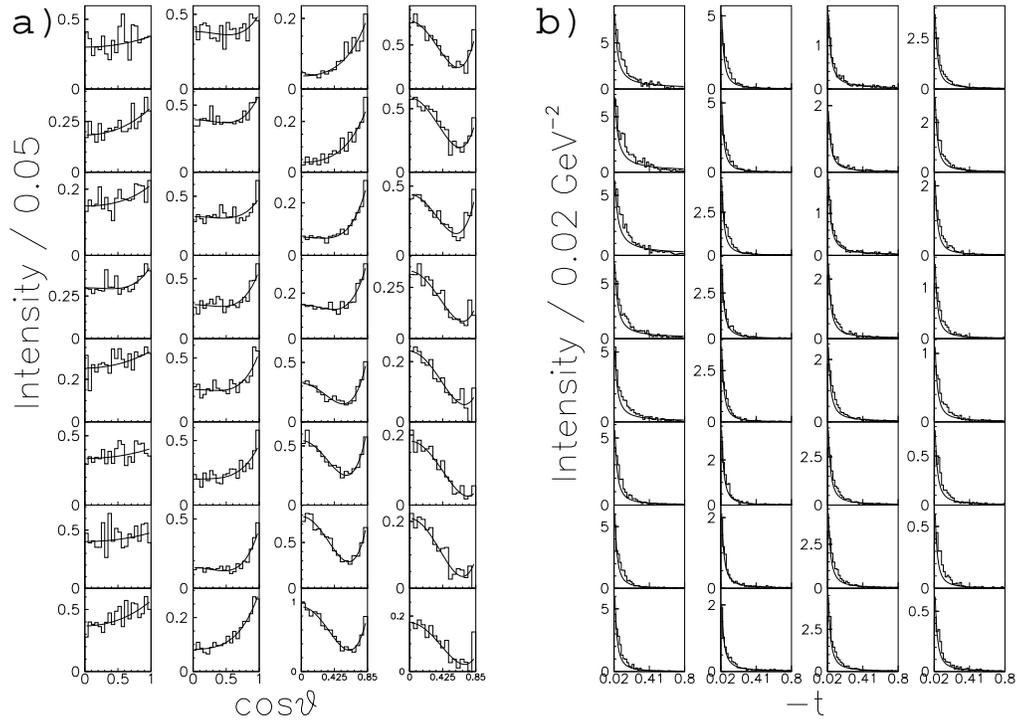}}
 \caption{ 	a) cos$\theta$ and b) $t$ distributions for 
each 40 MeV mass bin between 0.36 and 
1.64 MeV of  $\pi^0\pi^0$  mass. 
Solid lines show results of the fits. Details are in the text
}
  \label{fig2}
\end{figure}

	cos$\theta$ and $t$ distributions are shown in 
Fig. 2 a) and b) respectively, for each 40 MeV wide mass bin 
between 0.36 and 1.64 MeV of  $\pi^0\pi^0$  mass. 
The partial waves obtained 
are shown in Fig.3 a) and b) for the S and D waves, respectively. 
Breit-Wigner parameters are obtained above 1 GeV for the S and D waves, 
as follows; $M = 1278 \pm 5$ 
MeV and $\Gamma = 197 \pm 8$ MeV for $f_0(1370)$ and 
$M = 1286 \pm 57$ MeV and $\Gamma = 161 \pm 14$ 
MeV for $f_2(1270)$. 
Solid lines in Fig. c) and d) show the results of the fits. 
A fourth power  
polynomial background is used for the S wave.

	The S wave  $\pi^0\pi^0$  scattering amplitudes can be written 
$|A_S|^2 \sim {\rm sin}^2(\delta_S^0-\delta_S^2)$  below $K\bar K$ threshold
where  $\delta_S^0$ and  $\delta_S^2$ are 
the S wave scattering phase shifts of I = 0 and I = 2,
respectively.
We use a hard core type for $\delta_S^2$, as 
$\delta_S^2 = -r_c | {\bf q}_1 | = -r_c \sqrt{m_{\pi\pi}^2/4 - m_\pi^2}$, 
where $r_c$ is the hard 
core radius. 
The parameter $r_c$ has been obtained\cite{rf4} from $\pi^+\pi^-$ 
data\cite{rf5} so far. 
Fig.4 a) shows the normalized S-wave amplitude squared below 1 GeV.
The S-wave 
 $\pi^0\pi^0$  phase shifts obtained below $K\bar K$ threshold are 
shown in Fig.4 b) by solid squares. 
The results are consistent with $\pi^+\pi^-$ phase shift 
data below 650 MeV, though they appear  
somewhat higher than those above 650 MeV. The results are  
consistent with those of the down flat solution obtained in the reanalysis performed 
recently by Kaminski et al.\cite{rf6} on the CERN-Cracow Munich 
polarization data. 

\begin{figure}[t]
  \epsfysize=9. cm
 \centerline{\epsffile{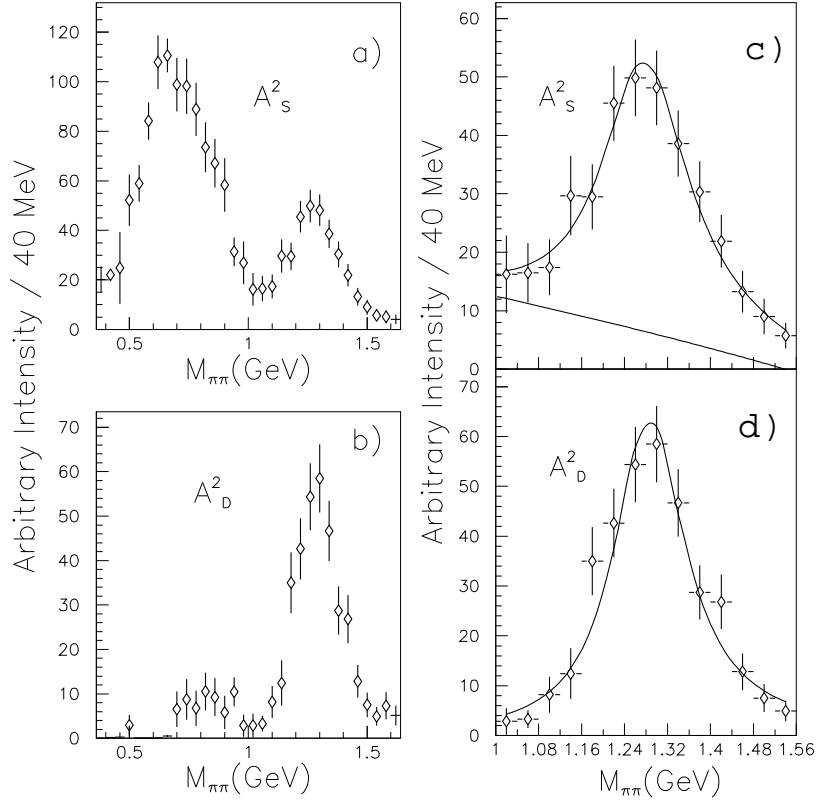}}
 \caption{Results of the partial wave analysis. a) for S wave, 
b) for D wave.\ \ 
c), d) Breit-Wigner fit above 1 GeV for S and D waves, respectively.
Solid lines are results of the fits.
}
  \label{fig3}
\end{figure}

	The  $\pi^0\pi^0$  phase shift data are analyzed by the 
 Interfering Amplitude (IA) method\cite{rf7}. 
A hard core is used for the negative background. 
$f_0(980)$ and $\sigma$ are the  contributing resonant 
states. 
The fit is shown by the solid line in Fig.4 c). The fitted 
parameters of BW for the lower mass resonance, $\sigma$ are as follows; 
$M_\sigma = 588 \pm 12$ 
MeV, $\Gamma_\sigma = 281 \pm 25$ MeV and 
$r_c = 2.76 \pm 0.15$GeV$^{-1}$. The $\chi^2/n_{d.o.f}$ value is 20.4/12. 
These values are in good agreement with those which we have obtained in 
our reanalysis\cite{rf4,rf7} for $\pi^+\pi^-$ phase shift data. 
We checked also the case with no negative background (without 
hard core, $r_c = 0$). The $\chi^2/n_{d.o.f}$ becomes worse:
85.0/13. The BW parameters 
deviate from those with the hard core 
as follows; $M_{"\sigma "} = 890 \pm 16$ MeV and 
$\Gamma_{"\sigma "} = 618 \pm 51$ MeV. The results is 
shown by the dotted line in Fig. 4 c).

\begin{figure}[t]
  \epsfysize=7.5 cm
 \centerline{\epsffile{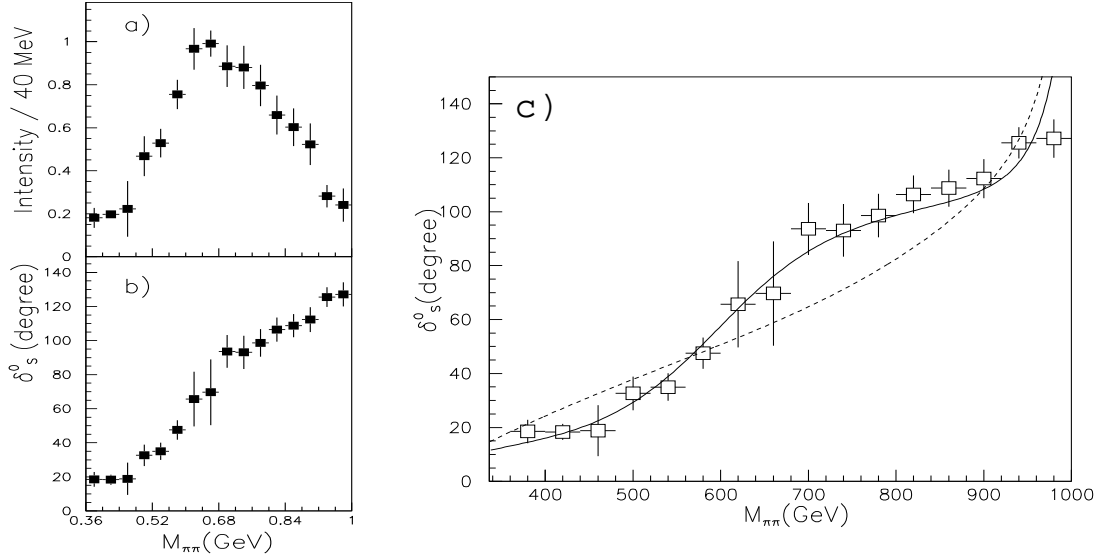}}
 \caption{ a) Normalized intensity of S wave amplitude squared below
   1 GeV, b) S wave $\pi^0\pi^0$ scattering phase shift $\delta_S^0$
   below $K\bar K$ threshold and c) the result of fitting by the IA
   method with the hard core (solid line) and without it.
   (dotted line). 
}
  \label{fig4}
\end{figure}

     The difference of I = 0 and I= 2 S wave phase shift, 
$\delta_S^0-\delta_S^2$ at the neutral kaon mass is 
related to the parameters of the CP violation in the 
neutral K decay. 
We obtain a value, $\delta_S^0 -\delta_S^2 = 42.5 \pm 3^\circ$,
in agreement with the previous result,   
40.6$\pm$3$^\circ$\cite{rf8}.

\end{document}